# On the Derivation of One- and Two-Dimensional Equations for Layered Elastic Beams and Plates with Interface Slips

Yilin Qu[1,2] and Jiashi Yang[3]

[1]School of Marine Science and Technology, Northwestern Polytechnical University, Xi'an, Shaanxi 710072, China

[2]Unmanned Vehicle Innovation Center, Ningbo Institute of NPU, Ningbo, Zhejiang 315048, China

[3]Department of Mechanical and Materials Engineering, University of Nebraska-Lincoln, Lincoln, NE 68588, USA, jyang1@unl.edu

**Abstract**

Mindlin's systematic procedure of power series expansion for deriving one- and two-dimensional equations of elastic beams and plates is extended to layered beams and plates with interface slips by adding a step function term to the power series expansion. A two-layer beam is used as an example. In addition to the usual one-dimensional equations for extension, bending and shear deformation, an equation associated with the interface slip also results from the procedure.

## 1. Introduction

Multilayered beams and plates are common structures in many engineering fields. In some cases, the bonding at the interfaces among different layers is not very strong and relative displacements between neighboring layers may occur. Such interfaces may be called weak or imperfectly bonded interfaces. Layered structures with imperfectly bonded interfaces have been under sustained studies for a long time, e.g., [1-8]. More references can be found in review articles such as [9]. One- and two dimensional equations for these structures can be obtained in the conventional manner by studying differential elements of the structures, or constructed by joining separate equations of individual layers using interface conditions [7,8].

Mindlin initiated a general procedure [10,11] of deriving one- and two-dimensional equations for beams and plates using power series expansions in the variational formulation of the theory of elasticity. In this paper, we generalize Mindlin's procedure to beams with interface slips by adding a step function term in the power series expansion. Then the rest of the derivation is straightforward.

## 2. Three-Dimensional Equations

Let the displacement vector, stress tensor and strain tensor be denoted by **u**, **T** and **S**. The strain and displacement are related by

$$S_{ij} = (u_{i,j} + u_{j,i})/2 . \tag{1}$$

The stress-strain relation may be written as

$$U(\mathbf{S}) = \frac{1}{2} c_{ijkl} S_{ij} S_{kl}, \quad T_{ij} = \frac{\partial U}{\partial S_{ij}} = c_{ijkl} S_{kl}, \tag{2}$$

where $U$ is the strain energy density and **c** is the elastic stiffness. For isotropic materials,

Young's modulus $E$, Poisson's ration $v$ and shear modulus $G$ are often used and

$$S_{ij} = \frac{1}{E}[(1+v)T_{ij} - vT_{kk}\delta_{ij}], \quad G = \frac{E}{2(1+v)}. \tag{3}$$

For a finite body occupying a region $V$ whose boundary surface is $S$ with an outward unit normal **n**, the equation of motion and the traction boundary condition are the stationary conditions of the following variational functional

$$\Pi(\mathbf{u}) = \int_{t_0}^{t_1} dt \int_V [\frac{1}{2}\rho \dot{u}_i \dot{u}_i - U(\mathbf{S}) + f_i u_i] dV + \int_{t_0}^{t_1} dt \int_S t_i u_i dS \tag{4}$$

whose first variation is

$$\delta\Pi = \int_{t_0}^{t_1} dt \int_V (T_{ji,j} + f_i - \rho \ddot{u}_i)\delta u_i dV - \int_{t_0}^{t_1} dt \int_S (n_j T_{ji} - t_i)\delta u_i dS, \tag{5}$$

where $\rho$ is the mass density, **f** is the body force, and **t** is the surface traction.

### 3. Derivation of One-Dimensional Equations for Beams

Consider the two-layer beam in Fig. 1. The dimensions and material properties of the two layers are the same.

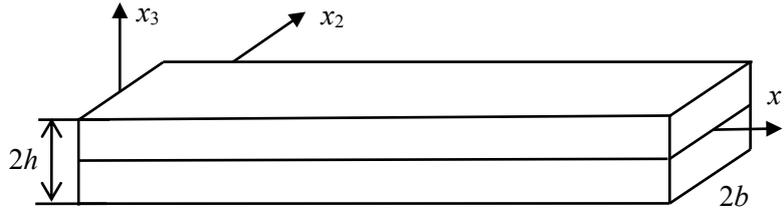

Fig. 1. A beam with two identical layers and coordinate system

For extension and bending in the $(x_1, x_3)$ plane, we approximate the relevant displacements by the following lower-order terms of power series expansions in $x_3$, with a step function term in $u_1$ to describe the interface slip:

$$\begin{aligned} u_3(\mathbf{x},t) &\cong u_3^{(0)}(x_1,t), \\ u_1(\mathbf{x},t) &\cong u_1^{(0)}(x_1,t) + x_3 u_1^{(1)}(x_1,t) + u(x_3) u_1^S(x_1,t) \\ &= \begin{cases} u_1^{(0)} + u_1^{(1)} x_3 + u_1^S, & 0 < x_3 < h, \\ u_1^{(0)} + u_1^{(1)} x_3 - u_1^S, & -h < x_3 < 0, \end{cases} \end{aligned} \tag{6}$$

where $u$ is the step function defined by

$$u(x_3) = \begin{cases} 1 & 0 < x_3 < h, \\ -1 & -h < x_3 < 0. \end{cases} \tag{7}$$

The interface shear elastic constant is denoted by $K$. At the interface where $x_3=0$, we have [12]:

$$T_{31}(0^+) = T_{31}(0^-) = K[u_1(0^+) - u_1(0^-)]. \tag{8}$$

The strain energy associated with the interface is given by

$$\int_0^L \int_{-b}^b \frac{1}{2} K[u_1(0^+) - u_1(0^-)]^2 dx_2 dx_1, \qquad (9)$$

which modifies (4). We want to obtain one-dimensional equation for bending described by $u_3^{(0)}$, extension by $u_1^{(0)}$, shear deformation by $u_1^{(1)}$ and interface slip by $u_1^S$. Substituting (6) and (9) into (4), using integration by parts, we obtain the following equations of motion and possible boundary conditions from the stationary conditions of (4). For $0 < x_1 < L$, we have

$$\begin{aligned}
T_{11,1}^{(0)} + f_1^{(0)} + 2b(t_1' + t_1'') &= \rho^{(0)} \ddot{u}_1^{(0)}, \\
T_{13,1}^{(0)} + f_3^{(0)} + 2b(t_3' + t_3'') &= \rho^{(0)} \ddot{u}_3^{(0)}, \\
T_{11,1}^{(1)} - T_{13}^{(0)} + f_1^{(1)} + 2bh(t_1' - t_1'') &= \rho^{(2)} \ddot{u}_1^{(1)} + (\rho'^{(1)} - \rho''^{(1)}) \ddot{u}_1^S, \\
T_{11,1}'^{(0)} - T_{11,1}''^{(0)} - 8bKu_1^S + (f_1'^{(0)} - f_1''^{(0)}) + 2b(t_1' - t_1'') \\
&= \rho^{(0)} \ddot{u}_1^S + (\rho'^{(1)} - \rho''^{(1)}) \ddot{u}_1^{(1)}.
\end{aligned} \qquad (10)$$

At $x_1 = 0$ or $L$, one may prescribe

$$\begin{aligned}
& T_{11}^{(0)} \quad \text{or} \quad u_1^{(0)}, \\
& T_{11}^{(1)} \quad \text{or} \quad u_1^{(1)}, \\
& T_{13}^{(0)} \quad \text{or} \quad u_3^{(0)}, \\
& T_{11}'^{(0)} - T_{11}''^{(0)} \quad \text{or} \quad u_1^S.
\end{aligned} \qquad (11)$$

In (10) and (11), the resultants over a cross section of the beam are defined by

$$\begin{aligned}
T_{11}^{(0)} &= 4bhE u_{1,1}^{(0)}, \\
T_{13}^{(0)} &= 4bhG\kappa^2 (u_{3,1}^{(0)} + u_1^{(1)}), \\
T_{11}^{(1)} &= -2bh^2 E u_{1,1}^S + \frac{4bh^2}{3} E u_{1,1}^{(1)}, \\
T_{11}'^{(0)} &= 2bhE(u_{1,1}^{(0)} - u_{1,1}^S) + bh^2 E u_{1,1}^{(1)}, \\
T_{11}''^{(0)} &= 2bhE(u_{1,1}^{(0)} + u_{1,1}^S) - bh^2 E u_{1,1}^{(1)},
\end{aligned} \qquad (12)$$

where $\kappa$ is a shear correction factor [10,11] artificially introduced to improve the accuracy of the equations. The plate inertias of various orders are defined by

$$\begin{aligned}
\rho^{(n)} &= 2b \int_{-h}^h \rho x_3^n dx_3 = \rho'^{(n)} + \rho''^{(n)}, \\
\rho'^{(n)} &= 2b \int_0^h \rho x_3^n dx_3, \quad \rho''^{(n)} = 2b \int_{-h}^0 \rho x_3^n dx_3, \\
n &= 0, 1, 2.
\end{aligned} \qquad (13)$$

A prime us for the upper layer. A double prime is for the lower layer. The plate loads from the body force **f** are

$$\begin{aligned}
f_i^{(n)} &= 2b \int_{-h}^h (f_i x_3^n) dx_3 = f_i'^{(n)} + f_i''^{(n)}, \\
f_i'^{(n)} &= 2b \int_0^h (f_i x_3^n) dx_3, \quad f_i''^{(n)} = 2b \int_{-h}^0 (f_i x_3^n) dx_3.
\end{aligned} \qquad (14)$$

$\mathbf{t}'$ and $\mathbf{t}''$ are the surface tractions at the top and bottom of the beam. Substituting (12) into (10), we obtain the following four equations for $u_1^{(0)}$, $u_3^{(0)}$, $u_1^{(1)}$ and $u_1^S$:

$$4bhEu_{1,11}^{(0)} + f_1^{(0)} + 2b(t_1'+t_1'') = \rho^{(0)}\ddot{u}_1^{(0)} \tag{15}$$

$$4bhG\kappa^2(u_{3,11}^{(0)} + u_{1,1}^{(1)}) + f_3^{(0)} + 2b(t_3'+t_3'') = \rho^{(0)}\ddot{u}_3^{(0)},$$

$$-2bh^2 Eu_{1,11}^S + \frac{4bh^2}{3}Eu_{1,11}^{(1)} - 4bhG\kappa^2(u_{3,1}^{(0)} + u_1^{(1)})$$
$$+ f_1^{(1)} + 2bh(t_1'-t_1'') = (\rho'^{(1)} - \rho''^{(1)})\ddot{u}_1^S + \rho^{(2)}\ddot{u}_1^{(1)}, \tag{16}$$
$$-4bhEu_{1,1}^S + 2bh^2 Eu_{1,1}^{(1)} - 8bKu_1^S$$
$$+ (f_1'^{(0)} - f_1''^{(0)}) + 2b(t_1'+t_1'') = \rho^{(0)}\ddot{u}_1^S + (\rho'^{(1)} - \rho''^{(1)})\ddot{u}_1^{(1)}.$$

## 4. Discussions

In the special case of a perfectly bonded interface with $K=\infty$, (15) and (16) reduce to

$$4bhEu_{1,11}^{(0)} + f_1^{(0)} + 2b(t_1'+t_1'') = \rho^{(0)}\ddot{u}_1^{(0)} \tag{17}$$

$$4bhG\kappa^2(u_{3,11}^{(0)} + u_{1,1}^{(1)}) + f_3^{(0)} + 2b(t_3'+t_3'') = \rho^{(0)}\ddot{u}_3^{(0)},$$

$$\frac{4bh^2}{3}Eu_{1,11}^{(1)} - 4bhG\kappa^2(u_{3,1}^{(0)} + u_1^{(1)}) + f_1^{(1)} + 2bh(t_1'-t_1'') = \rho^{(2)}\ddot{u}_1^{(1)}, \tag{18}$$

$$u_1^S = 0.$$

(18)$_{2,3}$ are the two equations of the well-known Timoshenko theory for a beam in bending with shear deformation.

Thickness modes are independent of $x_1$. For a free plate without body and surface loads, from (15) and (16), the thickness modes are governed by

$$0 = \rho^{(0)}\ddot{u}_1^{(0)} \tag{19}$$

$$0 = \rho^{(0)}\ddot{u}_3^{(0)},$$
$$-4bhG\kappa^2 u_1^{(1)} = (\rho'^{(1)} - \rho''^{(1)})\ddot{u}_1^S + \rho^{(2)}\ddot{u}_1^{(1)}, \tag{20}$$
$$-8bKu_1^S = \rho^{(0)}\ddot{u}_1^S + (\rho'^{(1)} - \rho''^{(1)})\ddot{u}_1^{(1)}.$$

(20)$_{2,3}$ show that there are two resonance frequencies for coupled thickness shear described by $u_1^{(1)}$ and interface slip by $u_1^S$. If the coupling is neglected, (20)$_{2,3}$ produce the following frequencies:

$$\omega^2 = \frac{4bhG\kappa^2}{\rho^{(2)}} \quad \text{for} \quad u_1^{(1)},$$
$$\omega^2 = \frac{8bK}{\rho^{(0)}} \quad \text{for} \quad u_1^S. \tag{21}$$

For layered elastic plates with interface slips, similar to (6), the displacement field may be approximated by

$$u_3(\mathbf{x},t) \cong u_3^{(0)}(x_1,t),$$
$$u_a(\mathbf{x},t) \cong u_a^{(0)}(x_1,t) + x_3 u_a^{(1)}(x_1,t) + u(x_3) u_a^S(x_1,t), \qquad (22)$$
$$a = 1, 2.$$

## 5. Conclusions

One- and two-dimensional equations for layered beams and plates with imperfectly bonded interfaces can derived systematically using Mindlin's power series procedure with the addition of step function terms for the description of the interface slips. The interface slips lead to additional equations. The procedure can be further generalized to the case when interface stretch is present.